\documentclass[letterpaper]{article}
\usepackage{aaai}
\usepackage{times}
\usepackage{helvet}
\usepackage{courier}\usepackage[utf8]{inputenc} 
\usepackage[T1]{fontenc}    
\usepackage{hyperref}       
\usepackage{url}            
\usepackage{booktabs}       
\usepackage{amsfonts}       
\usepackage{nicefrac}       
\usepackage{microtype}      
\usepackage{xcolor}         
\usepackage{comment}
\usepackage{graphicx}
\usepackage{enumitem}
\setlist{nolistsep,leftmargin=*}
\usepackage{natbib}[square,numbers]

\begin{document}
%

\title{Generating Synthetic Data in a Secure Federated General Adversarial Networks for a Consortium of Health Registries}
\author{Narasimha Raghavan Veeraragavan and Jan Franz Nygård\\
Department of Registry Informatics\\
Cancer Registry of Norway\\
Oslo, Norway\\
}
\maketitle
\begin{abstract}
\begin{quote}
In this work, we review the architecture design of existing federated General Adversarial Networks (GAN) solutions and highlight the security and trust-related weaknesses in the existing designs. We then describe how these weaknesses make existing designs unsuitable for the requirements needed for a consortium of health registries working towards generating synthetic datasets for research purposes. 
Moreover, we propose how these weaknesses can be addressed with our novel architecture solution. Our novel architecture solution combines several building blocks to generate synthetic data in a decentralised setting. Consortium blockchains, secure multi-party computations, and homomorphic encryption are the core building blocks of our proposed architecture solution to address the weaknesses in the existing design of federated GANs. Finally, we discuss our proposed solution's advantages and future research directions.

\end{quote}
\end{abstract}
\section{Introduction}
\noindent Health registries worldwide are established to collect data and monitor incidence, prevalence, and survival changes over time. Furthermore, health registries conduct, promote, and provide the basis for research. Synthetic data are precious when real data are expensive, scarce or simply unavailable due to a lack of consent from patients. Depending on the research’s objective, synthetic data can be used as a reasonable proxy or augment the real data to speed up the research pace~\cite{goncalves}. 

Even though there are several techniques available to generate synthetic datasets and the field has been researched for more than three decades, General Adversarial Networks (GANs) based methods have shown exciting results to create fake data that balances the utility and privacy~\cite{gan}. 

Health registries in a consortium may want to augment the real data stored locally with synthetic data generated based on the probability distributions of the real data located in other health registries. To this end, federated GAN-based solutions are attractive to the consortium of health registries. 

The existing federated GAN-based solutions do not have trust built into the design, as discussed in Section~\ref{sec:challenges}. To this end, we propose a novel architecture solution framework to generate synthetic data for a consortium of health registries. The framework uses the following technologies as building blocks: a) federated GANs~\cite{FedGAN} b) consortium blockchain~\cite{fabric} c) Shamir secret sharing algorithm~\cite{shamir} and d) homomorphic encryption~\cite{homomorphic}. 

The key contributions of the paper can be summarised as follows: 
\begin{itemize}
\itemsep0em
    \item Identified the trust-based challenges in the existing design of federated GAN solutions.
    \item Identified the requirements for a consortium of health registries to implement a federated GAN solution. 
    \item A novel architecture solution that can withstand stronger threat models to generate synthetic data for each member of a consortium of health registries. 
    \item An informal discussion about how the proposed architecture satisfies the requirements towards implementing a federated GAN solution for a consortium of health registries. 
\end{itemize}

\section{Background}
\subsection{Consortium Blockchains}
Consortium blockchains are meant to record transactions between the controlled consortium members on the immutable ledger stored on each member of the consortium. These transactions are responsible for tracking the state changes of an asset. State changes of an asset happen through the smart contracts that implement the business logic. An example of a consortium blockchain is Hyperledger Fabric~\cite{fabric}. 

Privacy is achieved in the consortium blockchain, such as Fabric, through channels and private data collections. A consortium can have multiple channels with a different subset of members. Transactions and smart contracts are isolated per channel, and members of the same channel hold an identical copy of the ledger (an immutable key-value store) and a state database (and a mutable database). The private data collection feature is used to enable the members of a channel to act as witnesses of the transactions rather than revealing all the details of the transactions between the transacting members of the channel. 

Recent advances in consortium blockchain technology enable processing 41,670 transactions per second with 20ms response time~\cite{bidl}. Hence, low latency and high throughput transactions are achievable with the consortium blockchain technology.

\section{Challenges in the designs of Federated GANs}
~\label{sec:challenges}
A vital element of the GAN training process is tight integration between the generator and discriminator since the generator network is trained based on the outputs from the discriminator, and the discriminator requires the output from the generator to train the discriminator's model. In a centralised GAN-based design, the integration between the generator and discriminator is more robust since the generator and discriminator are usually located in a single machine and controlled by a single administrative entity. However, federated GANs based on client/server architecture style~\cite{MD-GAN} have loose coupling between the global generator and local discriminator since the global generator and local discriminators are located on different machines and controlled by various administration entities.   

Even though secure communication channels (for example, using PKI techniques) can be established between the server and all the clients to prevent man-in-the-middle attacks on the communication channel, there is still a fundamental question related to the trust exist between a centralised coordinator/intermediary/server and multiple clients due to the different administration units involved in the entire federated GAN training process. 

The question of trust in the federated GAN can be divided into the following general questions: 
\begin{itemize}
\itemsep0em
    \item Can clients fully trust the operations contributed by the server/intermediary towards training federated GAN?  
    \item Can the server fully trust the operations contributed by the client towards training federated GAN? 
    \item Can each client trust the operations contributed by other clients towards training federated GAN? 
\end{itemize}

In each variant of client/server architecture style-based federated GAN, the above questions transform into concrete questions relevant to implementing the variant. For example, in the case of MD-GAN~\cite{MD-GAN} where a server hosts a single generator and each client hosts a discriminator under iid data source settings, the trust-related questions are as follows: 
\begin{itemize}

    \item Did all the targeted discriminators hosted on all clients use the appropriate outputs generated by the centralised generator as part of each iteration of the training processes of discriminators' networks? 
    \item Did the centralised generator use all the received and targeted discriminators' outputs to update its weights and compute the generator's gradients for each training iteration of the generator? 
    \item Did all the targeted discriminators receive the correct outputs from the centralised generator for each iteration of discriminators' training?  
    \item Did the swapping happen correctly between all the discriminators at the correct periods? 
\end{itemize}

An alternative solution to the client/server architecture style for federated GAN is peer-to-peer gossiping-based approaches, as discussed in~\cite{g-gan}. However, federated GANs based on peer-to-peer solutions still have a question of trust among the peers contributing to the training operations of federated GANs. In other words, can each peer fully trust the processes contributed by other peers towards training the federated GAN?

For example, \cite{g-gan} proposes that each machine/peer/client sends its generator and discriminator model parameters to another randomly selected neighbour after $K$ local iterations. And the chosen randomly neighbour averages the received parameters with the local generator and discriminator parameters before running a new learning step of $K$ iterations. In this case, the critical trust-related questions are as follows:  

\begin{itemize}

\item Did the peer who received the model parameters use the received model parameters for average computation?
\item Did all the peers send the correct model parameters for each $k$ iteration to other peers in the network?
\item Did all the peers who received the model parameters compute the average between the local and the received parameters?
\end{itemize}

These questions manifest themselves into several attack surfaces for conducting model and data poisoning attacks. The root causes of these questions are due to the fact that the existing federated GAN designs lacked accountability, auditability, data confidentiality, and decentralized trust. In our proposed design, we ensure that the before-mentioned properties are provided as an integral part of our system design.

\section{System Model}
In our proposed model, the health registries located in each country act as separate administration units and host patients’ medical records. Health registries do not share personal medical records with other health registries located in other countries due to the sensitive nature of the documents. However, research institutions and health registries form a consortium to facilitate research projects such as Nordcan~\cite{nordcan}, Decansec~\cite{decansec}, Vantage6~\cite{vantage6}, and Datashield~\cite{datashield}.

It would be very beneficial for a consortium of health registries to generate synthetic medical records based on their local medical records and medical records from other health registries to promote research.  
\subsection{Assumptions}
\begin{itemize}

\item Each health registry isolates the real data required for the training process in a separate database and has agreed on a common data format to be used for the training process in an offline fashion. 
\item The real data stored in the health registries are horizontally partitioned. 
\item Each registry's discriminator trains on the local data and updates the trained models based on predefined algorithms. 
\item Each registry runs the training on its data centres with a dedicated secure communication channel to interact with other registries' data centres.       
\item Each registry runs several training iterations on local data before merging with the global updates. 
\item All smart contracts are code-reviewed by all consortium members before being deployed by the network. 
\item Health registries use a Public Key Infrastructure-based system for authentication and authorisation purposes. 
\item Generator and discriminator model architectures used by the consortium members are homogeneous and agreed in an offline fashion. 
\end{itemize}
\subsection{Threat Model} 
\begin{itemize}
\itemsep0em
    \item A centralised server/intermediary can exhibit arbitrary/byzantine behaviour, which can lead to model poisoning attack and breach of data confidentiality 
    \item Peers in a decentralised solution can deny their actions leading to repudiation attacks. 
    \item Model and data poisoning attacks can happen due to manipulated inputs and outputs generated during the training process. 
\end{itemize}
\subsection{Requirements}
The following requirements are the properties that should be inherent as part of the federated GAN design to address trust-related questions and mitigate the risks due to model and data poisoning attacks.
\begin{itemize}
\itemsep0em
    \item \textbf{Accountability}: each registry should be accountable for the inputs used during the generator network training and discriminator network training in addition to the outputs generated by these networks. This accountability requirement in turn minimizes the risk due to model and data poisoning threats.  
    \item \textbf{Data confidentiality:} each registry's real-world data used as part of discriminator network training operations should not be revealed to other registries. Additionally, all the gradients and parameter vectors used as part of the training process of one registry should not be revealed to another registry. 
    \item \textbf{Auditability:} an external auditor can verify all registry transactions related to the training process that led to the final global generator model creation. 
    \item \textbf{Decentralized trust:} no registry can be individually trusted regarding their contribution to acting as a server.  
\end{itemize}

\section{Proposed scheme}
\begin{figure}[!htb]
\minipage{0.5\textwidth}
  \includegraphics[width=\linewidth]{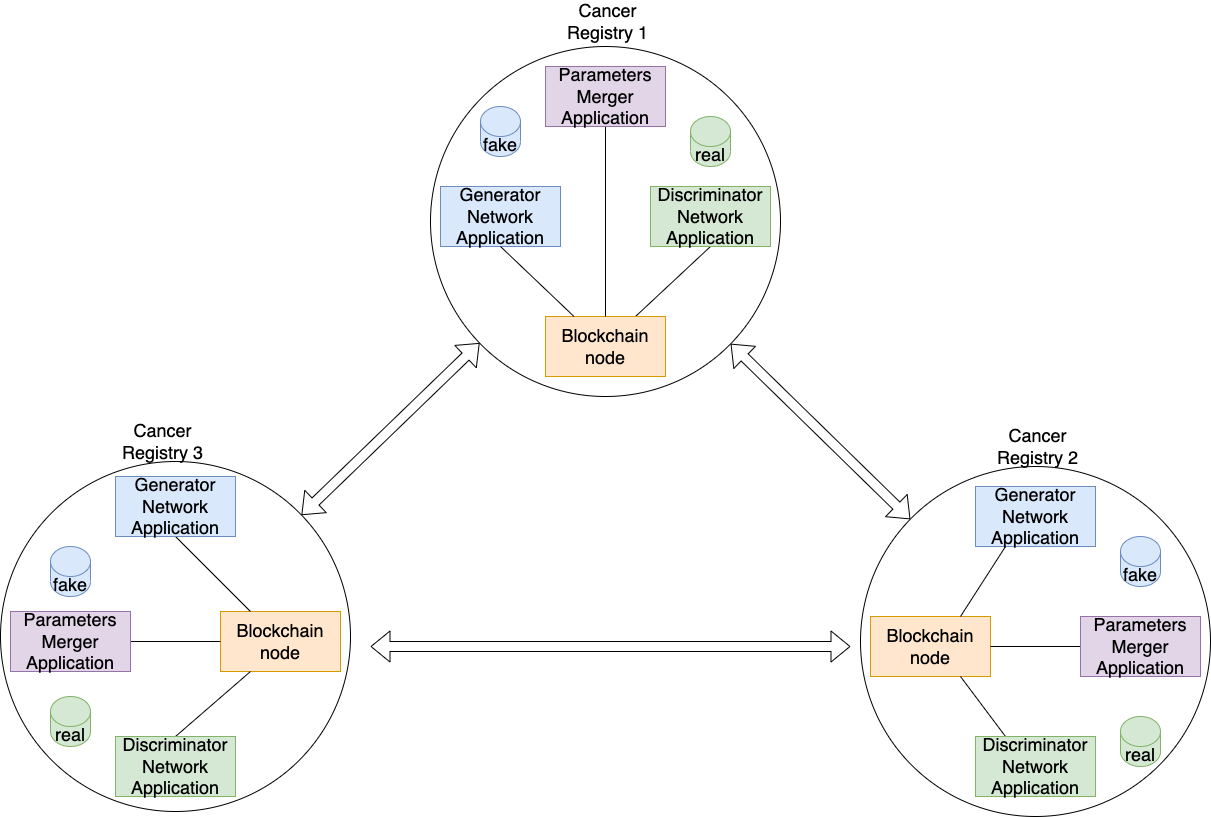}
  \caption{An Overview of a Consortium of health Registries with three members}\label{fig:overview}
\endminipage\hfill
\minipage{0.5\textwidth}%
  \includegraphics[width=\linewidth]{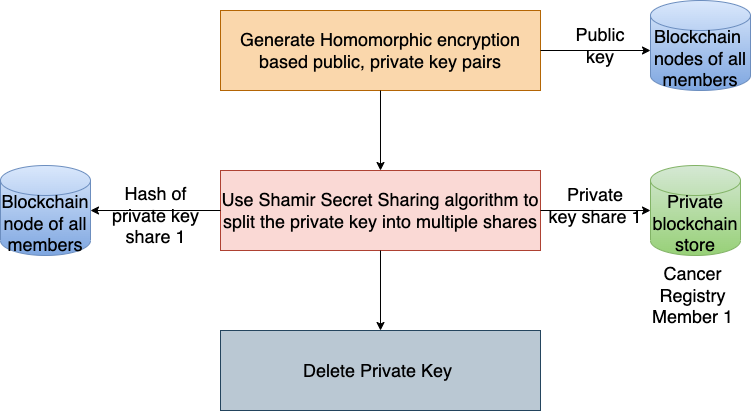}
  \caption{Initialization phase for a consortium of health registries }\label{fig:initialization}
\endminipage\hfill
\minipage{0.5\textwidth}
  \includegraphics[width=\linewidth]{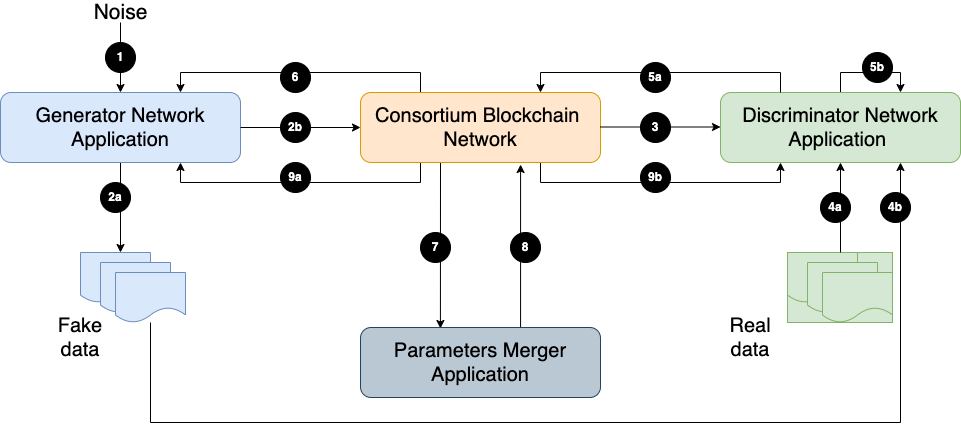}
  \caption{A proposed workflow for each member of a consortium}\label{fig:flow}
\endminipage

\end{figure}
We divide our proposed solution into four phases: a) initialization b) generator network application, c) discriminator network application d) parameters merger application. 
\subsection{Initialization} 
The initialisation phase is performed infrequently. The purpose of the initialisation phase is to create channels and deploy all the smart contracts essential for each channel to be operational. Our proposed scheme has two main channels: a) the key-generation and distribution channel and b) the GAN applications interaction channel. 

The key-generation channel has $|N|$ private data collections where $N$ represents the set of members of the consortium. The channel has smart contracts to generate a public and a private key pair based on homomorphic encryption. Homomorphic encryption-based keys are used to support averaging-based operations on encrypted model parameters. Furthermore, the channel has smart contracts to split the private key pairs. 

The generator and discriminator use the public key to record encrypted gradients and weights to the blockchain. The private key is not required during the training process; however it is needed to verify the training process as part of the auditability requirements. Considering the sensitivity of the private key, the private key should remain confidential. To this end, the private key is split into $|N|$ shares using the Shamir secret sharing algorithm. The private key shares are recorded in the corresponding private stores of the channel members. After recording the private key shares, the private key used for splitting is destroyed to ensure that no one member can decrypt the encrypted records from the consortium blockchain; the private key shares can be combined later to create the private key to satisfy the auditability requirement. 

In addition to the key generation channel, we create a channel for GAN applications to interact with the consortium blockchain. The purpose of the channel is to record the encrypted local model parameters of the generator and discriminator of each consortium member in parallel. Additionally, the channel can record the average model weights of the generator and discriminator computed by the merger application of each consortium member. 
In essence, the GAN application interactions channel has smart contracts that can record the encrypted model parameters and the average computation on the encrypted model weights. 

\subsection{Generator Network Application}
The generator in each member of the consortium executes the following steps: 
\begin{itemize} 
\itemsep0em
\item  The generator takes the noise from a random distribution as the input only for the first round of training (\textbf{flow 1} in Figure~\ref{fig:flow}). 
\item  The generator generates synthetic data and records the output of the synthetic data in the local filesystem and creates a hash of the synthetic data (\textbf{flow 2a} in Figure~\ref{fig:flow}). 
\item  The generator uses the public key to encrypt the weights and gradients of the model to the consortium blockchain network in addition to the hash of the synthetic data (\textbf{flow 2b} in Figure~\ref{fig:flow}). 
\item  The generator receives the notification event about the generator loss from the discriminator via the consortium blockchain. If the generator loss is unacceptable, the generator uses the generator loss output to update the generator model weights and gradients (\textbf{flow 6} in Figure~\ref{fig:flow}). Else, the generator loss is accepted, the generator training iteration is stopped, and the stop discriminator training event is recorded to stop the discriminator training. 
\item If the generator receives the notification event about the averaged weights, then the generator uses the averaged weights to update the model weights (\textbf{flow 9a} in Figure~\ref{fig:flow}). 

\end{itemize} 

An iteration for a generator network application is from the time to get an event notification from the consortium blockchain network until the time that event gets processed to produce the fake data. The event notification is either the generator loss output from the blockchain network (flow 6 in Figure~\ref{fig:flow}, or the averaged weight event (flow 9a in Figure~\ref{fig:flow}).  The training of the generator stops when the generator receives the acceptable generator loss.

\subsection{Discriminator Network Application} 
The discriminator in each member of the consortium executes the following steps: 
\begin{itemize} 
\itemsep0em
\item  As soon as the generator has recorded its transaction in the consortium blockchain network, the discriminator receives the notification event from the consortium blockchain network (\textbf{flow 3} in Figure~\ref{fig:flow}). If the notification event is stopped, the discriminator stops the training or proceeds to the next step. 
\item  The discriminator takes the samples from the real data (\textbf{flow 4a} in Figure~\ref{fig:flow}) in addition to the fake data samples from the local file system (\textbf{flow 4b} in Figure~\ref{fig:flow}). 
\item  The discriminator uses the hash of the fake data samples and checks if the computed hash is the same as the hash in the blockchain network; if the same, proceed forward with the classification task and compute the discriminator and generator losses. 
\item  The discriminator uses the public key to encrypt the weights and gradients of the model to the consortium blockchain network in addition to the discriminator and generator loss. Additionally, the hash of the real data used for the classification is also recorded (\textbf{flow 5a} in Figure~\ref{fig:flow}). 
\item  The discriminator uses the discriminator loss to update its model weights and gradients (\textbf{flow 5b} in Figure~\ref{fig:flow}). 
\item If the discriminator receives the notification event about the averaged weights, then the discriminator uses the averaged weights to update the model weights (\textbf{flow 9b} in Figure~\ref{fig:flow}). 
\end{itemize} 

An iteration for a discriminator network application is from the time to get an event notification from the consortium blockchain network. Until the time that event gets processed to produce the discriminator and generator losses. If the event notification is stopped, the discriminator does not proceed with the training process. 
\subsection{Parameters Merger application} 

The parameters merger application in each member of the consortium executes the following steps: 
\begin{itemize}
\itemsep0em
    \item The consortium blockchain notifies the parameter merger application after every $K$ iteration to compute the average along with the set of encrypted weights of generators and discriminators (\textbf{flow 7} in Figure~\ref{fig:flow}) that are recorded in the consortium blockchain by other registries. 
    \item The parameter merger application computes the average on the encrypted records and records it in the consortium blockchain (\textbf{flow 8} in Figure~\ref{fig:flow}). 
\end{itemize}

We expect at least $N$ where $N > 1$ occurrences of $K$ iterations to happen before any registry reaches its acceptable generator loss. Once the registry reaches its acceptable generator loss, the consortium blockchain network sends the stop notification event to the parameter merger application, which exits the application. 
\section{Discussion}
In this section, we discuss the design rationale of our solution and how our solution satisfies the proposed requirements. 

The Generator, Discriminator and Parameter Merger applications are chosen as client applications to a consortium blockchain network rather than treated as an integral part (smart contracts) of the consortium blockchain network. This loosely coupled design choice allows the consortium members to experiment with different GAN model architectures without disturbing the consortium blockchain setup. Additionally, the parameter merging algorithms are (either simple FedAvg or modified FedAvg) confined by the homomorphic computations.

Because the model architecture is homogeneous, the acceptable threshold for generator loss remains common across the consortium members. As soon as the allowable threshold is reached, the training stops. We want to explore more about model convergence in the near future experiments.

The workflow in the proposed architecture is designed so that all the inputs (except the real data) and outputs of each training iteration of the generator and discriminator are recorded in the blockchain. Due to the traceability property offered by the consortium blockchain, each local training iteration of the generator and discriminator can be traced back to its origin. Thus, each registry member becomes accountable for the inputs and outputs produced by their local GAN training processes. 

Additionally, the design satisfies the data confidentiality requirement by recording only the encrypted model parameters and a hash of the synthetic and real data. Furthermore, the keys to retrieve the encrypted records can only be obtained through the combination of private key shares, which requires explicit requests to each registry.       

An approved external auditor or a data protection authority can conduct a thorough audit of the training process of each registry by officially collecting the private key shares from registries. And the auditor or the authority can combine the collected key shares to create the master private key to decrypt the encrypted records retrieved from the blockchain—the immutability property of the blockchain guarantees that the encrypted records do not tamper. Thus, the requirement of auditability is satisfied by the proposed design. 

Furthermore, the proposed workflow design does not need any single server to coordinate the training process between the registries due to the use of the consortium blockchain network. The peer-to-peer nature of consortium blockchain guarantees coordination in a decentralised way with the help of gossip protocols. Hence, a single point of failure can be avoided.

\section{Related Work}

The existing works in the literature of federated GANs can be divided into two types: a) the client/server style and b) decentralised gossip-style based solutions. All federated GAN solutions that depend on client/server-based architecture suffer from a single point of failure. Additionally, the existing client/server solutions~\cite{MD-GAN}~\cite{FedGAN}~\cite{PerFED-GAN} does not work under the byzantine threat model.

Decentralised gossip-based federated GANs~\cite{g-gan} suffer repudiation attacks where the peers responsible for the coordinated training process can act maliciously and deny the action after performing it. There is no means to hold them accountable for the existing designs. 

To the best of our knowledge, none of the existing federated GAN solutions has the following four properties/features:  accountability, auditability, data confidentiality and decentralised trust against strong threat models. 
\section{Conclusion and Future Work}
In this paper, we have investigated introducing trust in federated GANs. The trust concept is captured with the help of properties required by the infrastructure and the infrastructure's ability to function in the presence of stronger threat models. To the best of our knowledge, the proposed solution makes the first attempt to introduce trust by design for generating synthetic data among a group of entities hosting distributed health registry data. 

In the near future, we plan to implement the proposed architecture in collaboration with suitable partners, identify the performance and dependability bottlenecks, and offer solutions to fix the identified bottlenecks. Furthermore, there are various attempts proposed in the literature to evaluate the quality of synthetic data in a centralised setup~\cite{pmlr}~\cite{goncalves}~\cite{sajjadi} and the tutorial session in ICML 2021~\cite{icml2021} also pointed out that defining metrics to assess the quality of synthetic data generated by generative models as a tricky/challenging problem. 

\bibliographystyle{aaai}
\bibliography{references}
\end{document}